\documentclass[10pt]{article}
\usepackage[utf8]{inputenc}
\usepackage[T1]{fontenc}
\usepackage{graphicx}
\usepackage{amsmath}
\usepackage{subcaption}
\usepackage{authblk}
\usepackage{hyperref}
\usepackage{geometry}
\title{Most Western African migrants remain local and travel short distances}

\author[1,*]{Irene Tafani}
\author[2]{Ola Ali}
\author[2]{Rafael Prieto-Curiel}
\author[1,3]{Massimo Riccaboni}

\affil[1]{IMT School for Advanced Studies, AxES, Lucca, Italy}
\affil[2]{Complexity Science Hub, Metternichgasse 8, 1030 Vienna, Austria}
\affil[3]{IUSS, Palazzo del Broletto, 27100 Pavia, Italy}

\affil[*]{\texttt{irene.tafani@imtlucca.it}}

\date{} % leave empty for no date

\begin{document}
\maketitle

\begin{abstract}
Migration patterns are complex and context-dependent, with the distances migrants travel varying greatly depending on socio-economic and demographic factors. While global migration studies often focus on Western countries, there is a crucial gap in our understanding of migration dynamics within the African continent, particularly in West Africa. Using data from over 60,000 individuals from eight West African countries, this study examines the determinants of migration distance in the region. Our analysis reveals a bimodal distribution of migration distances: while most migrants travel locally within a hundred km, a smaller yet significant portion undertakes long-distance journeys, often exceeding 3,000 km. Socio-economic factors such as employment status, marital status and level of education play a decisive role in determining migration distances. Unemployed migrants, for instance, travel substantially farther (1,467 km on average) than their employed counterparts (295 km). Furthermore, we find that conflict-induced migration is particularly variable, with migrants fleeing violence often undertaking longer and riskier journeys. Our findings highlight the importance of considering both local and long-distance migration in policy decisions and support systems, as well as the need for a comprehensive understanding of migration in non-Western contexts. This study contributes to the broader discourse on human mobility by providing new insights into migration patterns in Western Africa, which in turn has implications for global migration research and policy development.
\end{abstract}

\section{Introduction}

%%% migration important, distance not traveled euqally by everyone

Migration is a fundamental aspect of human history and development, of great importance both for the regions of origin and destination and for the migrants themselves. Social scientists have long been intrigued by migration, from Ravenstein's six laws of migration in 1885 to the current debates that dominate the media and generate different and often divergent points of view \cite{ravenstein1885laws}.

%%% what do we know about distances.... there are some unkowns

Theoretically, migration encompasses two dimensions: the spatial and the temporal. The temporal dimension refers to the length of time a person must live in a new place in order to be considered a migrant, while the spatial dimension refers to the distance a person must travel to be considered a migrant \cite{niedomysl2011migration}. Spatial distance has a negative impact on migration \cite{DistanceOnMigration, stouffer1940intervening, anderson2011gravity}. However, the determinants of the outward mobility friction in migration models could be related to the lack of information about distant opportunities or to the social and psychological costs of distance \cite{CostMigration, DistanceOnMigration, basu2022migration, Schwarz1973}. Thus, in general, most migrants tend to travel short distances \cite{tobler1995migration, bernard2017comparing, barbosa2018human}. Similar short distances for migration have been observed for internal migrants in Europe, where most migrations take place within 100 km \cite{halas2021revealing, karachurina2023migration}. Yet, the distance travelled by migrants is not uniform but varies widely due to different motivations and context-specific effects \cite{swedenmotives, worldmig}, and it may also differ depending on whether the migration is voluntary or part of a forced migration phenomenon. This distinction is particularly significant, as Sub-Saharan migrants represent both the largest and fastest-growing group of forced migrants and refugees \cite{annualreview}.
This variability is evident in Western Africa, where recent research has shown that migration intentions in response to weather shocks differ significantly across countries \cite{weather}. Understanding these differences is, therefore, crucial to comprehensively address the challenges and opportunities associated with migration. Existing research on migration distances highlights a complex interplay of socio-economic, and personal factors that influence how far individuals are willing or able to travel \cite{CHEN2008519}. For instance, while some migrants may relocate within their immediate region, others undertake long and often dangerous journeys across continents.

%%% why is distance relevant local and far away. (integration and return)

Distance is a critical factor in migration studies as it affects integration processes in host communities and the potential for return migration. Migrants travelling shorter distances may find it easier to maintain connections with their home communities, facilitating potential return. In contrast, those who migrate over long distances might face greater challenges in integration but also experience different socio-economic opportunities. For these reasons, the distance travelled can indicate the friction of long-distance journeys in different countries and thus reflect the extent of the search space considered in migration \cite{distance_larry}.  This is particularly critical in the case of forced long-distance migrations that take place due to a lack of migration opportunities in nearby regions. Although the literature on migration distance and its motives is limited\cite{niedomysl2011migration}, it has identified consistent patterns among Western countries. For example, it is well-established that long-distance migration is primarily driven by employment and educational opportunities \cite{swedenmotives,basu2022migration}. Family-related migration, on the other hand, varies depending on the context. Sometimes, it is associated with local moves, and other times with both local and long-distance migrations. %While this research project does not aim to focus on the distinction between long and short distances, it is crucial to raise awareness about this concept. Depending on their objectives, researchers may use cutoffs, administrative borders, or continuous measures of distance; the latter being used in this paper. Anyway, the scope of these studies is typically confined to intra-country migration.\\

%%% are these pattern equal or vary in africa?  impact of conflict in it. NO STUDIES IN AFRICA. 

Empirical studies usually focus on national migration movements and often deal with distances of up to a few hundred km \cite{niedomysl2014,blumenstock2023, Stillwell2016, Morrison2011}. In contrast, our study has the opportunity to analyze migrations of up to almost 6000 km, which gives us the chance to get an overview of an entire Western African region. This broader framework allows us to uncover patterns that would otherwise not be detectable if we only looked at migration within a country. Despite the expanding body of literature on migration patterns, a significant gap remains in our understanding of how these distances vary within the context of non-Western countries, such as Western Africa. Indeed, most research to date has focused on countries such as Great Britain \cite{Dixon2003UK, Thomas19UK, Champion1998}, US \cite{US}, Sweden \cite{niedomysl2011migration}, New Zealand \cite{Morrison2011} and Australia \cite{clarkaustralia}, mainly due to data availability, since official statistics fail to capture migration moves \cite{Akokpari2000}. To the best of our knowledge, while some studies have analyzed migration distance in relation to country-level characteristics such as SDI and income\cite{distancemobility}, none have yet explored the relationship between migration distance and individual migrant characteristics across the broader African continent.
However, research on migration in specific African countries does exist. For example, studies focusing on Ghana \cite{Beals1967} have produced findings that align with ours. Furthermore, other researchers have investigated the drivers of emigration in various African nations \cite{vanDalen2005} and analyzed migration intentions, accounting for differences in gender, marital status \cite{Gubhaju2009} and education \cite{Mittelmeier2021}. Previous research has emphasized that such studies are highly context-specific \cite{swedenmotives}. While certain findings from studies conducted in the Western world may be applicable to other culturally similar regions, we cannot make the same assumptions for Africa. The continent's unique socio-political environment necessitates a distinct approach to understanding migration patterns and their influencing factors. In Africa, migration patterns and distances travelled are influenced by unique factors, including regional conflicts, geophysical constraints, economic opportunities, and historical ties \cite{Naude2008Conflict}. Conflicts, in particular, add another layer of complexity with respect to other regions, driving people to migrate both within and across borders \cite{Crisp2006,oh2024emergent}. There is a notable lack of comprehensive studies focusing on how these factors play out in West Africa, a region characterized by diverse migration flows. Filling this gap is critical to gaining a nuanced understanding of migration dynamics on the continent, which is essential for developing effective policies and support systems tailored to the African context.

%%% How did we fill this gap. africa + contrast to deterring effect of distance

To address this gap, our study investigates the distribution of migration distances among people interviewed in Western Africa, analyzing how these distances vary based on different demographic and socio-economic features.
Given the limitations of gravity models in capturing temporal dynamics\cite{gravitybad} and predicting migration in response to shocks, as well as their reliance on population data, which is often unavailable, particularly in regions close to the Sahara desert, we turn to advanced predictive methodologies such as Random Forest and XGBoost models. These tree-based algorithms enable us to better account for the complex, non-linear relationships between migration trends and their driving factors and thus to identify the most influential factors driving migration patterns. In addition, by incorporating permutation importance testing, we can assess the relative importance of various factors in shaping migration distances and validate our findings. 
Our results show that migration distances in Western Africa are notably correlated by factors such as employment status, nationality, travel motives, marital status, age and educational level. In contrast, gender has a lesser impact. People displaced by conflict often exhibit highly variable travel patterns, moving locally one year and both locally and long distances the next, likely influenced by the conflict's intensity. In contrast, those who move for family reasons show consistent behaviour over the three years studied. They generally travel to the immediate vicinity with an average distance of around 100 km. In addition, employment status is identified as the most important factor in predicting the desired travel distance and shows considerable variability between the different categories. For example, the average travel distance for employed persons is 295 km, while for unemployed persons it is 1,467 km. Overall, migration tends to be a process of short distances, even if certain groups, such as unemployed migrants, may have to cross borders.

\section*{Results}
%%%% which data is used ?
{
In our study, we analyzed data collected by the International Organization for Migration in Western African countries (Senegal, Guinea, Gambia, Mali, Burkina Faso, Niger, Nigeria and Chad), focusing on interviews with migrants at critical transit points \cite{IOM_citation}. We pre-processed and obtained more than 60,000 observations over the course of three years (2021-2023) (Fig. \ref{fig:overall}). Although it is
 a convenience sample, this large dataset allowed us to investigate the distances travelled from the migrants' places of residence to their desired destinations. We calculate the distance traveled by the migrants using the geographical coordinates of their origin. These coordinates are obtained by manual labeling and from the Africapolis dataset, which covers all urban areas of the continent \cite{Africapolis}. Not surprisingly, our results show that most migration occurs over short distances, as is generally expected in gravity migration models. Specifically, the data show that individuals travel an average of 1,063 km from their place of origin to their destination (roughly equivalent to the distance between South and North Nigeria), with half of the respondents moving within 555 km. Having confirmed that our distribution is statistically different from random (Supplementary Materials Fig. 2), the observed distance distribution also shows a clear bimodal pattern (Fig. \ref{fig:overall}). There is a significant peak in the 0-100 km range and a second peak around 3,500 km. This indicates that the majority of West Africans either remain in the region or, if they move, often travel considerable distances of over 3,000 km. This bimodal distribution pattern has also been observed in other contexts where populations are geographically unevenly distributed (e.g. the east and west coasts of the USA) \cite{cerina2014}. Our results are consistent with this, as our data span areas that include the sparsely populated Sahara Desert.
}
%%%% figure africa total dist map+ hist
{
\begin{figure}[ht]
    \centering 
    \includegraphics[width=1.0\textwidth]{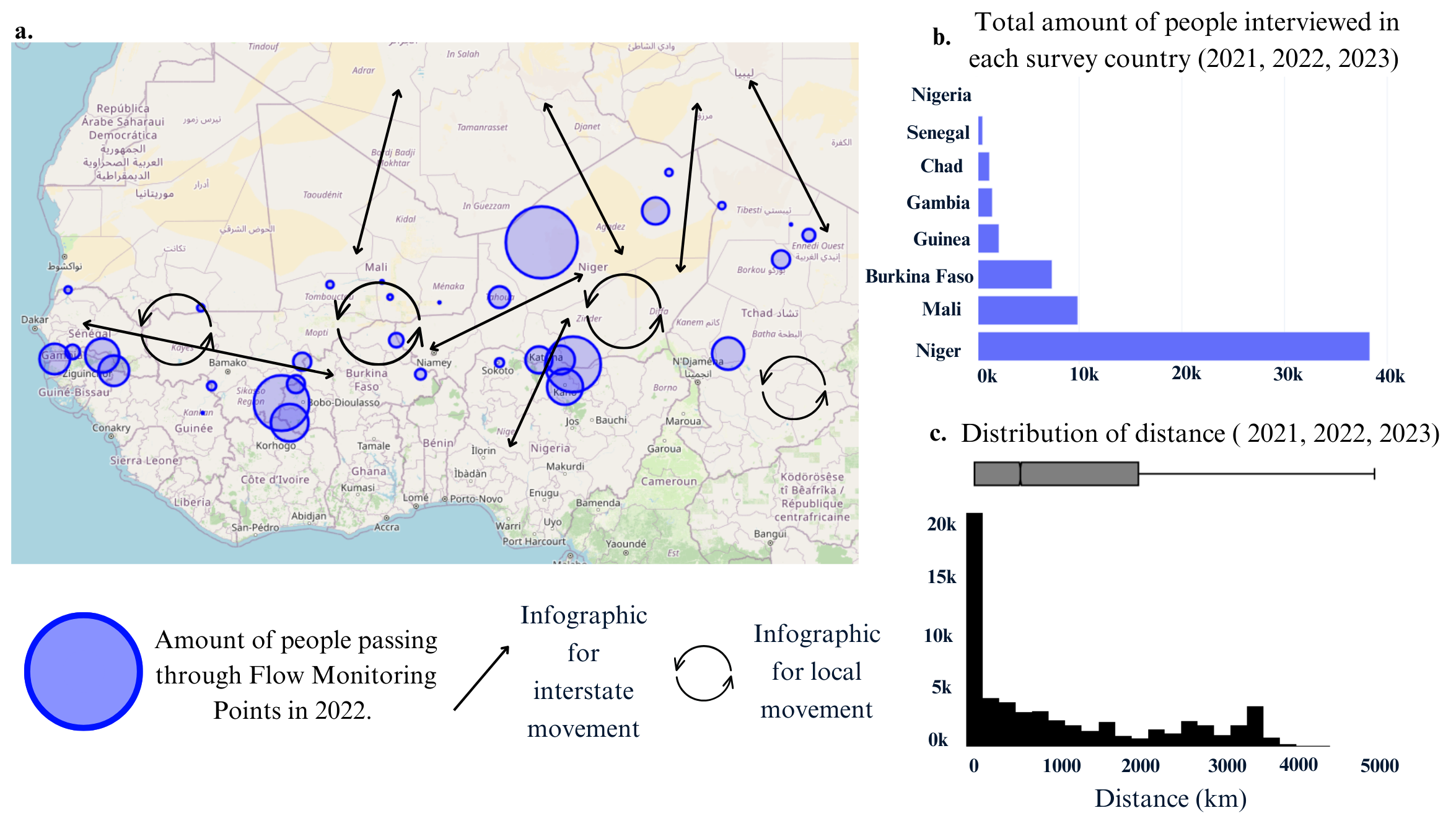} 
    \caption{Overview of the area of interest and the migration phenomenon under investigation. (a) Number of people traveling through the Flow Monitoring Points in 2022, together with the main movement flows within the Western African territory. (b) Number of people interviewed in each country. (c) The distance of all people surveyed from 2021 to 2023 shows a bimodal distribution with a prominent peak and a smaller, more distant peak.} 
    \label{fig:overall} 
\end{figure}
}

%%%% yet, reasons vary the distance travelled
{
The distance traveled by migrants varies greatly within the different socio-demographic groups and shows either a concentration of movement near the place of origin or both a local and a distant peak of migration.
}

%%%% 
{
When we consider demographic and socioeconomic characteristics, including employment status, marital status, reasons for migration, nationality, age, education and gender of respondents, we find that we can distinguish between those who travel locally and those who travel further afield ( Supplementary Material Fig. 4).
% We find that migrants travelling for economic reasons exhibit a bimodal distance distribution, indicating they tend to migrate either locally or to distant locations (Fig \ref{fig:distribution}). This bimodal pattern is also observed among single and unemployed migrants and those with lower educational levels, suggesting a similar split between local and faraway destinations.
% In contrast, the  distance distribution of migrants travelling for family reunification, educational purposes, and those with higher educational levels shows a single local peak, indicating a concentration of migrations within shorter distances.\\
}

%%% conclusion (?)
{
% These findings provide insights into the migration behaviors and trends of individuals based on their personal characteristics. They reveal the distances that each respondent in West and Central Africa is willing to travel, offering critical data for understanding migration dynamics in the region.\\
}

%%%% figure densities
{
% \begin{figure}[htbp]
%     \centering
%     % First subfigure
%     \begin{subfigure}[b]{0.45\textwidth}
%         \centering
%         \includegraphics[width=\textwidth]{images/density_gender.png}
%         \caption{Economic network}
%         \label{fig:sub1}
%     \end{subfigure}
%     \hfill
%     % Second subfigure
%     \begin{subfigure}[b]{0.45\textwidth}
%         \centering
%         \includegraphics[width=\textwidth]{images/density_marital.png}
%         \caption{Conflict and violence network}
%         \label{fig:sub4}
%     \end{subfigure}
%     \caption{Overall caption describing the figure as a whole.}
%     \label{fig:map_net}
% \end{figure}
}

\subsection{Who travels locally?}
%% what is local? 
{
In studying the Western African region, the concept of `local movement' has a unique significance compared to studies conducted in higher income countries. Although the area is comparable in size to Australia, the dynamics of local movement in Western Africa differ considerably. This is mainly due to the large differences in the size of nations and population density in West Africa. The Gambia, for example, stretches for around 300 km along its namesake river, while countries such as Nigeria, Mali, Niger and Chad stretch for over 1,000 km. Despite this wide range of distances in the region, we find that 35\% of people in Western Africa generally travel within 100 km. Furthermore, 38\% of people in our sample travel within the country, with an average inter-country distance of 49 km when we consider countries with at least one Flow Monitoring Point. This `home bias' effect, where people prefer to travel shorter distances within their home country rather than crossing borders, highlights an often neglected aspect of mobility in West Africa that needs to be taken into account when analyzing local movements.
}

%% results 
{
People who tend to stay close to home can be identified by certain personal characteristics. Starting with education level, we find that people with higher education and religious education tend to move within a range of 679 and 805 km. When looking at employment status, it is noticeable that retirees and employees (excluding the self-employed) move an average of 295 km. In terms of marital status, widowed and married individuals are the most likely to move locally, with an average range of 489 to 654 km, followed by divorced people, who travel an average distance of 905 km. Regarding the reasons for moving, family and education-related motivations are given as the main reason for the change of location, with an average distance of 404 km and 358 km respectively. Taking into account the nationality of the migrants and focusing only on those nationalities whose country has at least one Flow Monitoring Point in its territory, it can be seen that people from Nigeria and Burkina Faso tend to travel more locally, with an average of 677 and 655 km respectively. Looking at the age of respondents, people over the age of 33 tend to travel within a range of 861 km. With increasing age, the range of movement tends to decrease. Finally, our results show no significant difference in the distance traveled between men and women, with an average distance of 1,080 km for men and 1,089 km for women.
}

%%% image of the differences of distance distribution throughout the years and explainantion  ????
{
}

\subsection{Who travels far?}

%%%%
{
Those who travel far can be identified by characteristics that complement those mentioned above. If we look at the level of education, we see that people with vocational training travel an average of 1,728 km, which is roughly the distance from Athens (Greece) to Paris (France), while those with secondary education travel an average of 1,253 km, which is the distance from Abidjan (Côte d'Ivoire) to Bamako (Mali). Looking at employment status, the unemployed travel the furthest from their place of residence with an average distance of 1,467 km, followed by the self-employed, who travel an average of 837 km.
In terms of family status, singles are,as expected, the ones who travel the furthest away, followed by divorcees with an average distance of 1,514 km and 905 km respectively. Interestingly, in terms of reasons for moving, the factors that lead to longer distances are conflicts (1,269 km) and economic reasons (1,175 km), followed by climate-related reasons (1,061 km).
}

%%%%
{
The nationalities that tend to travel the longest distances are primarily Nigerians (2,052 km), followed by Guineans (1,335 km) and Senegalese (1,204 km), all with an average of over 1,000 km. The age group that moves the most is the youngest, aged between 18 and 22, who travel an average of 1,408 km.
}

\subsection{Using artificial intelligence to predict migration distance}

%%% much information!
{
We have given both a general and a detailed overview of how people in the Western African region tend to move. However, to fully understand the phenomenon of migration in this region, it is crucial to know which individual characteristics are most important to distinguish between those who are likely to move long distances and those who prefer shorter distances. These characteristics can be used to better predict expected migration flows and provide timely humanitarian assistance.
}

%%% which one is the most important?
{
To determine which personal characteristics of migrants were most relevant in predicting distance, we used two non-parametric tree-based predictors: Random Forest and XGBoost. Both methods provide a ranking of predictors from most to least important. After permutation tests of importance, we obtained identical rankings for both regressors, with nationality ranking first, followed by employment status and reasons for moving. The fact that these models conveyed the same ranking enhances the robustness of our findings. These results are also consistent with other studies on migration drivers in the sub-Saharan region \cite{ejq004, Naude2008Conflict}. Consistently with our preliminary descriptive analyses, gender ranked last, suggesting minimal explanatory power in predicting distance, even when conditioning on travelling in a group or alone.
}

%%% saying we also controlled for GDP
{
In addition to personal characteristics, we also included economic data, in particular the GDP per capita of the region of origin and destination, as predictors of migration distance. In line with the gravity approach, the inclusion of GDP data, especially GDP of the destination region, significantly increased the explanatory power of our models and led to an R-squared of up to 0.95. It is noteworthy that the GDP of the destination and the place of origin rank first in the importance of the features (Fig. 8 Supplementary Material). However, we decided to exclude the GDP data from our final analysis for several important reasons. First, the GDP of the actual destination is only known after the movement is completed and cannot be used to predict future movements. In addition, socio-economic instability in the region leads to significant fluctuations and uncertainties in GDP estimates, making them less reliable for understanding migration dynamics. Second, GDP data in many areas are not updated frequently enough to capture rapid changes due to conflicts, natural disasters or economic shocks that are prevalent in the study region. Third, the GDP of the target region conveys too much information that may overshadow the contributions of other socio-demographic factors that are central to our study. To maintain consistency and focus, we limited our analysis to information obtained directly from the surveys to ensure that our findings were based on data reflecting migrants' lived experiences rather than extrapolated economic indicators. After removing the GDP data, the R-squared values were 0.40 for XGBoost and 0.36 for Random Forest.}

%%% we focus on unemployment and conflict
{
At this stage, we focus on analyzing two main categories in predicting travel distance: employment status and reasons for travel. We found that the distribution of travel distance varies by status within each category. Therefore, we examined the two statuses associated with the greatest travel distances: Conflict and Unemployment (Fig. \ref{fig:conf_unemp}).
}

%%% figure conflit unemp 
{
\begin{figure}[ht]
    \centering \includegraphics[width=1\textwidth]{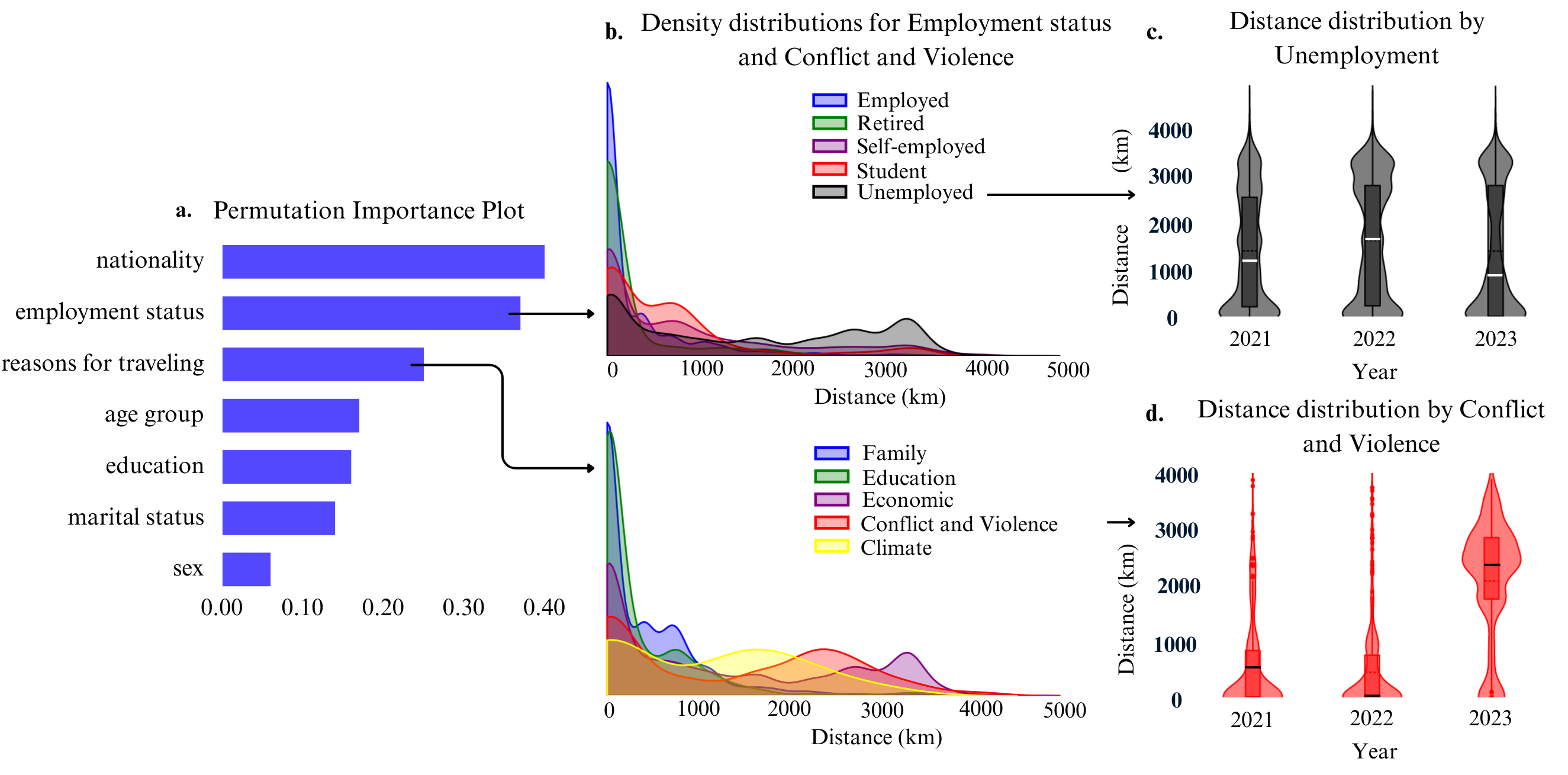} 
    \caption{The reasons for conflict-induced travel fluctuate more strongly compared to the steady trend in unemployment-induced migration. (a) The Permutation Importance Plot shows the most influential factors in predicting migration distances in West Africa. (b) shows the density distributions for employment status and reasons for travel, highlighting that migrants travel farthest due to unemployment and conflict. (c) and (e) show the annual trends for these categories, with conflict-related movements being very unstable compared to the consistent pattern of unemployment-related travel. This instability persists when comparing the annual distribution of conflict to a baseline without conflict and violence (Fig. 5 Supplementary Materials).} 
    \label{fig:conf_unemp} 
\end{figure}
}

%%% conflict vs unemplyment and baseline
{
Our aim is to analyze the movements of people in the Western African region, focusing on how individual characteristics of migrants influence whether they move locally or over long distances. Through our analyzes, we found that these individual characteristics, for example unemployment, show a generally consistent distribution over time ( Fig. \ref{fig:conf_unemp}), with the notable exception of conflict-related movements. This consistency in the individual characteristics can be observed despite the changes in the Flow Monitoring Points over the three years studied. It should be noted that the increase in migration flows between Niger and North Africa (Fig.\ref{fig:conflict_maps}) was not accompanied by an increase in the number of interview stations staffed by field researchers (Fig. 1 Supplementary Material).
}

%%% maps of conflicts throught the years
{
\begin{figure}[ht]
    \centering \includegraphics[width=0.9\textwidth]{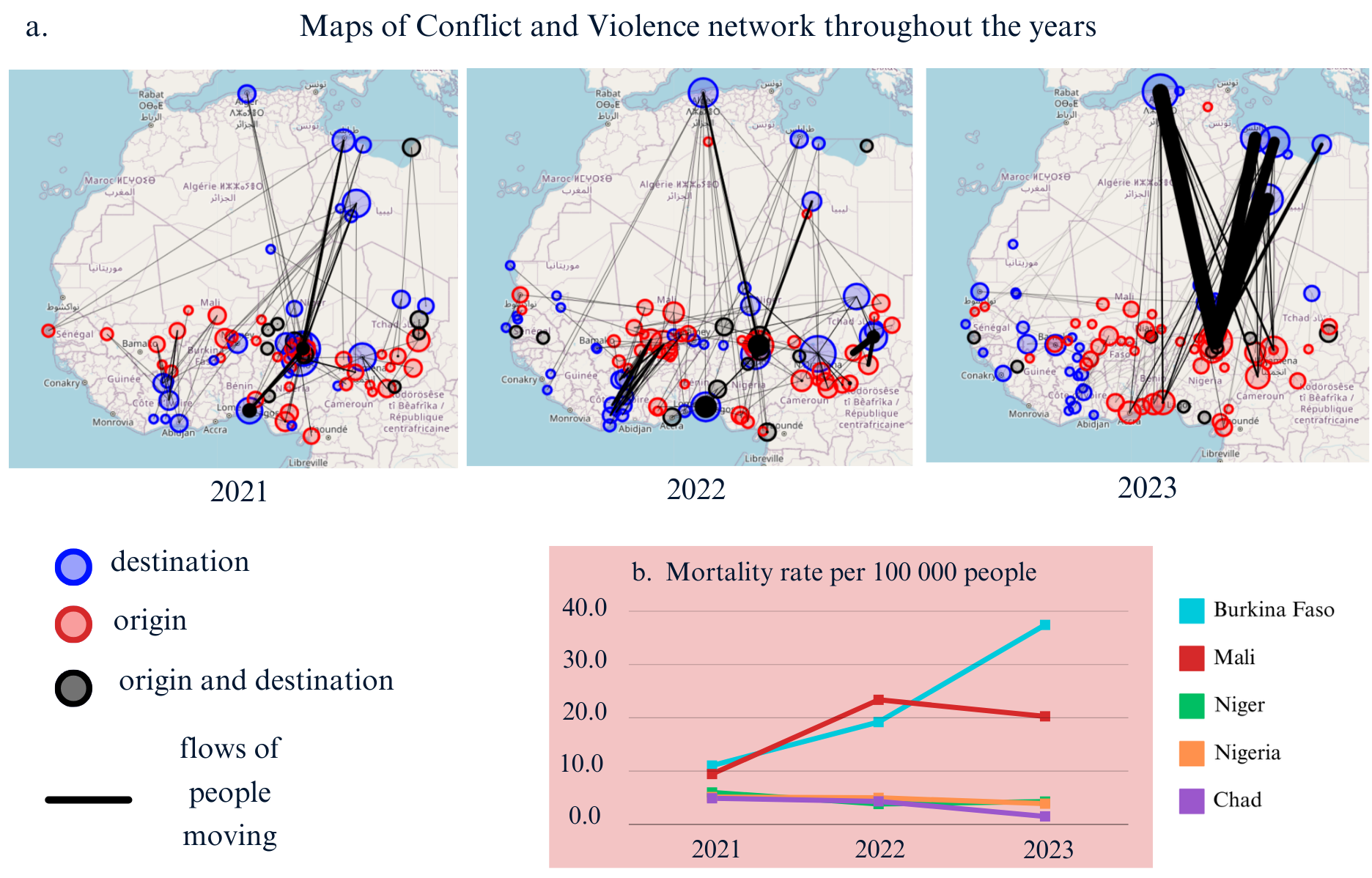} 
    \caption{Using the networks that represent the conflict movements over the three years, we can follow how the migration patterns have developed. In 2023, a year characterized by increased long-distance travel, regions that were previously destinations, such as Burkina Faso, Mali, Niger and Nigeria, have become regions of origin. Nigeria in particular is now the main source of migration across the Sahara to North Africa. This is consistent with data from the ACLED dataset, which shows that Nigeria has the highest number of fatalities, followed by Burkina Faso, where the number of fatalities jumped in 2023. Instability in the region is likely the reason for these longer migration routes.} 
    \label{fig:conflict_maps} 
\end{figure}
}

%%% proving the variability of conflict through multinomial...
{
The variability of migration distances due to conflict and violence is further supported by an additional analysis using multinomial logistic regression. This approach is consistent with previous studies in the literature and supports the conclusion that conflict-related movements are unpredictable and fluctuate more compared to other migration factors. This analysis examines the likelihood of migrants travelling along a span of 5,000 km of their place of residence, depending on various reasons: economic, family, education, climate and conflict. While the trends were stable in 2021 and 2022, there was a significant change in 2023. In this year, migrants fleeing conflict were the ones who traveled the furthest distances, in contrast to previous years where economic migrants dominated (Fig. \ref{fig:multinomial}). In addition, this analysis reveals another finding: contrary to some findings in the literature, people who migrate for educational reasons are more likely to prefer local moves to long-distance travel.
}

%%% figure multinomial
{
\begin{figure}[ht]
    \centering \includegraphics[width=0.9\textwidth]{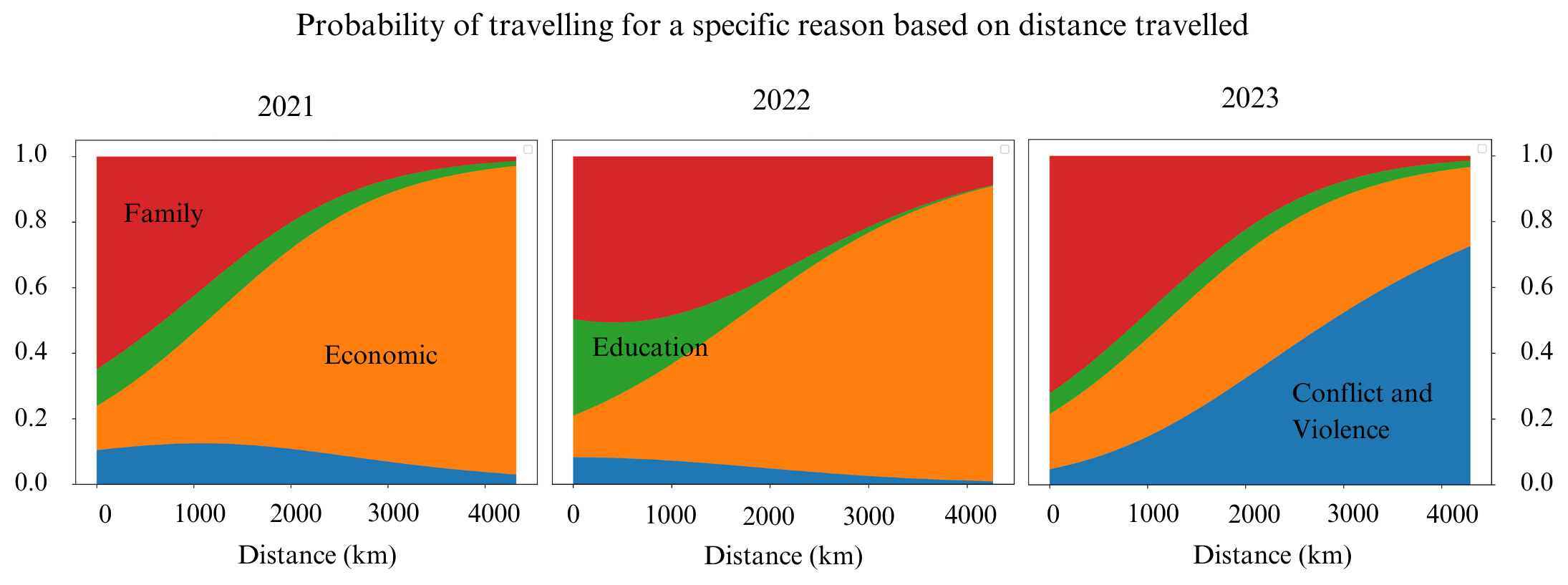} 
    \caption{The results of the multinomial regression analysis show that in 2021, the probability of encountering migrants traveling for family reasons decreases with increasing distance, while economic migration becomes more frequent. This trend remains broadly similar in 2022, although educational migration becomes more important locally. However, a notable change has occurred in 2023. At greater distances, migrants fleeing conflict are more common than economic migrants, which is a significant change from previous years.} 
    \label{fig:multinomial} 
\end{figure}
}

\section*{Discussion}

%%% the movements are local
{
Our findings reveal a complex landscape of migration patterns across Western Africa, characterized by significant variability in the distances traveled by different socio-demographic groups. While local migration predominates, with many people migrating within 100 km of their place of origin, our data also show a distinct bimodal distribution, suggesting that people who migrate beyond their immediate vicinity often undertake much longer journeys, typically over 3,000 km.
}

%%% who moves? how far?
{
In addition to this general trend of local movement, our analysis shows that certain demographic and socioeconomic characteristics play an important role in determining migration distances. For example, individuals with a higher level of education, stable employment or a certain marital status tend to migrate over relatively shorter distances, while unemployed or single migrants travel much further. Moreover, these patterns remain consistent over time, which is consistent with the expectations of gravity models that emphasize distance as a decisive factor in migration decisions. In these models, personal circumstances and economic conditions play an important role in determining how far people are willing or able to move \cite{greenwood}.
}

%%% but conflict is unstable
{
A notable exception to this stability, however, is conflict-induced migration. Unlike other factors, where migration distances remain relatively constant over time, conflict-related migration is highly variable. In 2023, for example, we observed a clear shift in which conflict-related long-distance migration was more pronounced than economic migration, which was more pronounced in previous years. This variability reflects the unpredictable and often rapidly changing nature of conflict zones, which can lead people to undertake much longer and more dangerous journeys in search of safety.
}

%%% policy
{
Understanding these patterns is critical to developing targeted interventions and policies that can better meet the needs of migrants in Western Africa. By identifying the factors that influence migration distances and recognizing the unique challenges posed by conflict-induced migration, this work provides valuable insights into the complex dynamics of migration in the region and offers a basis for more responsive and adaptive policymaking.

}
%%%% challengo to eurocentric view
{
Additionally, our findings challenge the common narrative in Western media and political discourse, which often portrays African migration as primarily driven by a desire to reach Europe. We offer a more nuanced perspective, illustrating that migration within Western Africa is largely local or regional, with many individuals migrating over short distances. This portrayal of Africa as a continent facing a migration 'crisis' towards Europe overlooks the diversity and complexity of migration patterns within the region. In reality, the majority of migration in Western Africa is driven by local factors, such as family dynamics, economic opportunities, and regional conflicts. Recognizing the local and regional dimensions of African migration is crucial, as it allows for a more accurate understanding of the motivations behind migration, moving beyond the simplification of migration as merely a Europe-bound phenomenon. Challenging these Eurocentric assumptions can help inform more effective and contextually relevant policies that take into account the diverse factors influencing migration within Africa itself.
}

%%% developments
{
Building on our findings, future research could include the development of predictive models for migration patterns using the demographic and socioeconomic factors identified in this study. Such models could improve predictive accuracy and help policy makers and humanitarian organizations to better anticipate and manage migration flows. In fact, machine learning offers promising opportunities to increase the effectiveness of humanitarian organizations in addressing migration-related challenges\cite{aiken2022}. By using these advanced data-driven techniques, it may be possible to improve the efficiency and fairness of resource allocation and ensure that aid reaches those most affected by migratory pressure and conflict.
Furthermore, examining the network properties of migration routes, especially in the context of conflict-related events, could reveal correlations that help predict migration patterns influenced by violence. Analyzing how these network characteristics interact with conflict dynamics could provide valuable insights into the movement of people fleeing conflict zones, allowing for more informed and targeted interventions to meet the needs of affected populations.
In general, a potential future development of this work could leverage network analysis techniques, focusing on the different reasons for travel. These networks (Supplementary Materials Fig. 4) exhibit distinct topological characteristics, which may reveal significant dynamics in understanding migration phenomena. 
}

\section*{Methods}

\subsection*{Data}
%% 4 IOM FMS data have been used so far only to study self-selection of refugees (Aksoy and Poutvaara,2019). https://www.econstor.eu/handle/10419/240201

%%IOM and Africapolis
{
The data used in this study refer to the years 2021, 2022 and 2023 and cover a total of eight countries: Senegal, Guinea, Gambia, Mali, Burkina Faso, Niger, Nigeria, and Chad. In 2022, a total of 34 Flow Monitoring Points were active, at which both the total inflow of passers-by and interviews were conducted. In comparison, there were 32 active Flow Monitoring Points in 2021, while the number decreased to 25 in 2023 (Tab. S1).}
%%%% 
{For our analyses, we focused on the municipal or provincial level of the migrants' place of residence and destination. Using these locations, we were able to calculate the distances between different points and derive the results. Determining the exact terrestrial locations of these places was a challenge due to their remote geographical location in desert areas and the different languages used for registration. Through meticulous effort, we mapped a total of 619 agglomerations, relying on both manual work and the Africapolis dataset\cite{Africapolis}. The latter provides standardized urbanization data for Africa, covering 54 countries with consistent data from 2015. It includes 7,496 small towns and intermediary cities and integrates various sources such as census data and satellite imagery. In our study, we used the unique codes assigned to different cities of different sizes and mapped them to our data. This allowed us to obtain reliable information on the coordinates of the different metropolitan areas in the countries considered.\\
}

%%%% talking about data in more detail.
{
In total, we collected 61,240 observations, each representing a migrant interviewed by researchers at active Flow Monitoring Points in specific countries and years, with modes of transportation varying between walking, car and bus. The activity of these Flow Monitoring Points varies from year to year as they are funded by external stakeholders whose interests and objectives may change according to changing needs and priorities. Despite this variability, these surveys are invaluable as they provide unique and comprehensive information, often collected in difficult and remote geographic locations. Because we have access to three years of data collection, we can perform detailed temporal analysis. However, these analyses are influenced by fluctuations in the number of data collection points. In addition, the limited sample size of climatic migrants (only 84 observations) further affects our results.
Nonetheless, the data remain an important resource for understanding migration patterns and trends, as well as insights into the socioeconomic and demographic profiles of migrants in different regions and time periods. To verify the presence of a signal in the empirical distribution of our data, we performed a permutation test, which confirms that the observed distribution differs significantly from a random distribution (Fig. 2 Supplementary Material).
}

%%% Acled data
{
Another data source we used is the Armed Conflict Location \& Event Data Project (ACLED) dataset to extract fatalities data \cite{acled}. The ACLED dataset is a comprehensive resource that provides detailed information on political violence and protests around the world. It contains data on different types of conflicts, such as battles, violence against civilians and riots, as well as associated fatalities. For this study, we focused specifically on the fatality data to understand the impact of conflict and violence on migration patterns. By analyzing the number of deaths in countries such as Nigeria, Burkina Faso, Mali and Niger, I was able to visually correlate the increase in deaths with the increase in long-distance migration flows in 2023. This dataset helped to highlight the regions with the highest rates of violence and provide context for the migration trends observed in the analysis.
}

%%% GDP data
{
Another source of information we use in our analysis is a comprehensive GDP per capita dataset \cite{gdpdata}, refined to the municipal level, covering 43,501 units from 1990 to 2022. This dataset updates previous versions that only included subnational data up to 2010. It was derived from the reported subnational GDP per capita data of 89 countries and 2,708 units, using advanced extrapolation and downscaling techniques from the field of machine learning \cite{kummu2022gdp_source}.
The dataset contains annual GDP per capita data at three administrative levels: national (Admin 0, with 237 units), provincial (Admin 1) and municipal (Admin 2). This granularity allowed us to accurately match our Admin 2 codes to the respective GDP figures, increasing the accuracy of our analysis.
}

\subsection{Tree based models}

{
Tree-based models such as Random Forest \cite{breiman2001random} and XGBoost\cite{xgboost} are powerful tools for predictive modeling and feature importance ranking. Random Forest is an ensemble learning method that builds multiple decision trees during training and outputs the mean prediction of each tree to improve prediction accuracy and control overfitting. The algorithm works by randomly selecting subsets of features and samples from the training data to build each tree, resulting in a ``forest'' of trees. The importance ranking in the Random Forest is derived from the average reduction in impurity (e.g. Gini impurity or variance in regression tasks) across all trees in the forest. The greater the reduction, the more important the characteristic is rated.

}
%%% why tree models?
{
 Tree-based models are particularly well-suited for this type of analysis because they can handle both continuous and categorical features, capture complex interactions between variables, and provide intuitive interpretations of feature importance. In the context of migration, where relationships between personal and economic factors and migration distance can be non-linear and involve intricate interactions, tree-based models are ideal for capturing these complexities. Additionally, compared to other models like linear regression, tree-based models do not require strict assumptions such as linearity or normality of residuals, making them more flexible in addressing real-world migration data.
}

{
XGBoost (Extreme Gradient Boosting) is another powerful ensemble learning technique that builds trees one after another, each trying to correct the errors of its predecessor. It uses gradient boosting, where trees are added to minimize a loss function. XGBoost includes regularization to prevent overfitting and uses a more sophisticated algorithm for tree construction. The importance of a feature in XGBoost can be measured by the overall gain, which is the improvement in accuracy that a feature brings to the branches in which it participates, or by the cover, which is the number of observations that relate to a feature. In addition, XGBoost offers other metrics, such as the frequency with which a feature is used to split data.
}
%%% permutation importance
{
An advanced technique for improving the robustness of feature importance rankings is Permutation Importance\cite{breiman2001random}. In this method, the values of each feature are shuffled and the change in the performance of the model is measured. By breaking the relationship between the feature and the target, the drop in performance indicates the importance of the feature. This technique provides a more reliable ranking by evaluating the actual impact of a feature on the model's predictions, rather than relying solely on how often or how favorable a feature is during tree partitioning. Permutation weighting has the added benefit of being model-agnostic and can be used to compare the importance rankings of different models such as Random Forest and XGBoost. This makes the rankings more consistent and robust and often leads to a convergence to an identical ranking, ensuring a more interpretable and trustworthy model evaluation, as was the case in our instance. In addition, we also used in-time cross-validation (Fig. 7 Supplementary Material ), where training is performed up to a certain point in time and then iteratively tested on a subset of observations.
}

\subsection{Multinomial Logistic Model}

{
In this study, a multinomial logistic regression is used to analyze the relationship between the reasons for moving and the distance travelled. A multinomial logistic model is suitable for modeling categorical outcomes with more than two levels and is thus ideal for our dependent variable, reasons for travel (family, economy, conflict, education), which consists of several different categories. In our case, the independent variable of interest is the distance traveled in continuous form. Therefore, the model can be viewed as a series of independent binary regressions

\[
\log \left( \frac{P(Y_i = j)}{P(Y_i = \text{baseline})} \right) = \beta_{0j} + \beta_{1j} \cdot \text{Distance}_i
\]

\noindent where \( Y_i \) denotes the reason for travel for the observation \( i \), \( \beta_{0j} \) and \( \beta_{1j} \) are the intercept and the coefficient for the distance traveled for the category \( j \) of the dependent variable.
This analysis allows us to estimate the likelihood of encountering a migrant based on one of the four categories of reasons for moving that depend on the distance traveled (Fig. \ref{fig:multinomial}).
This analysis allows us to estimate the likelihood of encountering a migrant based on one of the four categories of reasons for moving that depend on the distance traveled.
}

\bibliographystyle{plain}
%\bibliography{references.bib}

\section*{Acknowledgments}
The authors acknowledge partial support from the project “THE - Tuscany Health Ecosystem” (CUP: D63C22000400001), funded by the European Union - Next Generation EU program, in the context of the Italian National Recovery and Resilience Plan, Investment 1.5: Ecosystems of Innovation. 
The research was also funded by the Austrian Federal Ministry for Climate Action, Environment, Energy, Mobility, Innovation and Technology (2021-0.664.668) and the Austrian Federal Ministry of the Interior (2022-0.392.231). The authors also thank the International Organization for Migration (IOM) for providing access to the data that made this research possible.

\section*{Author contributions}
I.T., M.R. and R.P.C. conceived the experiment(s). I.T. conducted the experiment(s). All authors contributed to the interpretation of the results and reviewed the manuscript.

\section*{Competing interests}
The authors declare no competing interests.

\end{document}